# Resistive switching in ultra-thin $La_{0.7}Sr_{0.3}MnO_3$ / $SrRuO_3$ superlattices

S. Narayana Jammalamadaka[1,2], Johan Vanacken[1] and V. V. Moshchalkov[1]

[1] *INPAC – Institute for Nanoscale Physics and Chemistry, K.U. Leuven, Celestijnenlaan 200D, B–3001 Leuven, Belgium*

[2] *Magnetic Materials and Device Physics Laboratory, Department of Physics, Indian Institute of Technology Hyderabad, Hyderabad, Andhra Pradesh, PIN – 502 205, India.*

Corresponding author: surya@iith.ac.in

Superlattices may play an important role in next generation electronic and spintronic devices if the key-challenge of the reading and writing data can be solved. This challenge emerges from the coupling of low dimensional individual layers with macroscopic world. Here we report the study of the resistive switching characteristics of a of hybrid structure made out of a superlattice with ultrathin layers of two ferromagnetic metallic oxides, $La_{0.7}Sr_{0.3}MnO_3$ (LSMO) and $SrRuO_3$ (SRO). Bipolar resistive switching memory effects are measured on these LSMO/SRO superlattices, and the observed switching is explainable by ohmic and space charge-limited conduction laws. It is evident from the endurance characteristics that the on/off memory window of the cell is greater than 14, which indicates that this cell can reliably distinguish the stored information between high and low resistance states. The findings may pave a way to the construction of devices based on nonvolatile resistive memory effects.

Reaching closely to the limitations of Silicon-based technologies, research is driven by the quest of developing smaller, faster, cheaper and more capable electronic devices[1]. In this quest, devices which reproducibly switch their resistance state between a high resistance state and a low resistance state with respect to the voltage sweep (Resistive Random Access Memory devices, RRAM) have generated tremendous interest due to their low power consumption, high write/read speed, simple structure and high scalability characteristics[2-4]. The resistance switching can occur in different manners, *i.e.* a switching is called unipolar if the switching process does not depend on the polarity of voltage and current signal[2]. In contrast, if the switching depends on the polarity it is called bipolar resistive switching[2]. Several reports have shown such resistive switching characteristics on metal oxide films[5-8], organic films[9] and thin film based heterostructures[10]. It is however expectable that superlattice structures with few unit cells thick layers of metal oxides merge the low dimensionality ($\sim 10^{-9}$ m) of the individual layer thicknesses with the utility of large – scale films that can be handily linked to the real world[11]. Hence, one can establish a unique and potentially important stable resistance switching throughout transport measurement in a metal oxide superlattice structure.

Motivated by the above, here $La_{0.7}Sr_{0.3}MnO_3$ (LSMO) / $SrRuO_3$ (SRO) superlattices grown on SrTiO3(100) single crystals are studied for resistive switching and endurance characteristics. $La_{0.7}Sr_{0.3}MnO_3$ (LSMO)[12] and $SrRuO_3$ (SRO)[13] exhibit ferromagnetic order below their bulk Curie temperatures of 370 and 160 K, respectively. In these superlattices, the magnetically soft LSMO layers, are antiferromagnetically coupled with the magnetically hard SRO layers below 150 K, when all layers order ferromagnetically. In these superlattices by changing the magnetization state with a suitably applied magnetic field, exchange bias effects have been observed [14-16]. Having superlattice structure with 30 layers of LSMO and SRO allowed us to have more magnetic signal and hence significant exchange bias field, this may indeed allow us to control resistive switching behavior, which is of our future goal.

Keeping the above goal in mind, in this manuscript, we would like to demonstrate the resistive switching characteristics of LSMO/SRO superlattices which have not been studied until now. Precisely, in the present work we demonstrate: (a) resistive switching characteristics in a LSMO/SRO superlattice, with a validation of these results presenting various models; (b) endurance characteristics of the superlattice by changing the resistance state between on-state and off-state. Salient features of the present work are that a stable resistive switching is observed while sweeping the voltage and the current-voltage characteristics are well explained by space charge limited conduction (SCLC) model. Apart from that we also present results revealing a low–voltage (0.5 – 0.7 V) switching mechanism. Pertinent to the endurance characteristics, the on/off memory window of the cell is about 14, which indicates that this cell can be used to store information in the high- and low-resistance state.

Pulsed laser deposition technique, employing a KrF excimer laser, was used to fabricate a superlattice of $La_{0.7}Sr_{0.3}MnO_3/SrRuO_3$. Oxygen partial pressure and substrate temperature were 0.14 mbar and 650°C, respectively. In total 30 layers of $La_{0.7}Sr_{0.3}MnO_3$ and $SrRuO_3$ with thicknesses of 2.3 nm and 3.3 nm, respectively, were grown on $SrTiO_3$ (001) substrate. Details about the fabrication and the properties of such superlattices can be found in Ref. 17, 18. X–ray diffraction, atomic force microscopy, high resolution transmission electron microscope and high angle annular dark field scanning transmission electron microscopy (HAADF-STEM) were used to investigate the microstructure of the superlattice[17, 18]. HAADF-STEM micrograph of the $La_{0.7}Sr_{0.3}MnO_3$ / $SrRuO_3$ superlattice is shown in Figure 1. Misfit dislocations were found at the interfaces between the $La_{0.7}Sr_{0.3}MnO_3$ and $SrRuO_3$ layers. However, the interfacial atomic layers were affected by intermixing; both A-site (La/Sr) and B-site (Mn/Ru) cations intermix in 1-2 unit cells across the interface, as marked by the rectangles in Fig 1(b). The contacts (Fig. 1(a)) to the top layer were made by the wire bonder, the distance between two planar electrodes is 1 mm. Using Keithley 2400, two probe method was employed in order to measure I – V characteristics.

All the measurements were carried out at room temperature. Top SRO layer showed the resistance ~ KΩ. At this point it is worth quoting about the nature of SRO layer with different superlattice structures where essentially SRO layer behaved vividly in its bulk and ultra-thinfilm form. On top of that in general an important issue on using ultrathin films of metal oxides is the significant enhancement in the resistivity, which has been clearly manifested in SRO thin films[19-22].

Before the forming process, resistive switching at low voltages is not consistently observed in LSMO/SRO superlattice, between -1 V and +1 V. In order to make filaments between the electrodes, I – V characteristics were measured by sweeping voltage from 0 to 20 V with a current compliance of 10 mA. Essentially, the electro-forming process is necessary to get a stable and reproducible resistive switching behavior in the resistive switching devices. Compliance current of 10 mA is used to avoid the dielectric break down of the device during the forming process due to leakage current which may arise by the application of very high voltages. Essentially, during this process, current – limited electric breakdown is induced in the superlattice and subsequently, conductive ON state and a less conductive OFF states are streamlined. Immediately after the forming process, keeping the current compliance of 10 mA, the voltage was varied in the range of -1 and 1 V. Basically, stable switching of the voltages is observed in the low voltage limit. Starting with negative field sweep, initially the device is at low resistance state (LRS). During the voltage sweep rate, the device continued to be in the LRS up to - 0.5 V. At - 0.5 V, a sudden drop in the leakage current is evident as a consequence of the high resistance state (HRS). This process is called as RESET switching. Further increase in the voltage causes the device to stay at the same HRS up to – 1 V. While in the reverse run, the device stayed at HRS up to the positive 0.7 V and above which there is a sudden increase in the leakage current as a result of change in the resistance state to LRS. This process is called as SET process. Above 0.7 V, the leakage current is constant as a result of compliance limit. From positive 1 V, the device stayed in low resistance state up to 0 V. This kind of change in resistance state from LRS – HRS – LRS is a signature of bi – polar resistive switching[23-25]. This indeed means that in order to change the resistance state, voltage polarity needs to be changed. It is evident from the Fig. 2(a) that the resistive switching behavior is quite stable. Above procedure was repeated for 20 times to observe the stable RS. The arrows on the figure show the sequence of the applied voltage. Fig. 2(b) shows the same graph in logarithmic plot. As soon as we increased the compliance limit until 100 mA, we could see dielectric breakdown in the device. Hence, we fixed the compliance limit to 10 mA for all the devices.

To probe the current conduction mechanism of LSMO/SRO superlattices, I – V characteristics are analyzed with respect to space charge limited conduction (SCLC) mechanism[26, 27]. Logarithmic I – V plots for both the positive and negative regions are depicted in Fig. (3a, 3b).

Fig. 3(a) shows the log – log plot of such device. During the positive voltage sweeping, as we mentioned, the device is at high resistance state and changed its memory state to LRS. The initial region of the log – log plot is followed the linear ohmic behavior. In contrast, the fitting results of HRS are more complicated and exhibit different switching mechanism. Essentially, the charge transfer mechanism comprises of three parts, namely, ohmic region ($I \propto V$), the Child's law region ($I \propto V^2$), and the steep current increase region, which is in accordance with the classical space charge limited conduction (SCLC)[26, 27]. Fig. 3(b) shows the log – log plot pertinent to the negative voltage sweep region. In the negative voltage sweeping region, the behavior in LRS is similar to the positive voltage sweep region, however, the variation in the HRS is different. Such behavior in HRS region can be explained by a model called weak filamentary conduction channels[28]. According to this model, charge transport consists of two parts: weak filamentary channels current ($I_f$) and the current in bulk ($I_i$). In this model, when the temperature is below 250 K, $I_f$ dominates in HRS[28], however, in the present case, since the measurements were conducted at room temperature, current in HRS goes with the SCLC principle. Ultimately, the conduction behaviors in both the LRS and HRS also suggest that the high conductivity in on-state device should be a confined, filamentary effect rather than a homogenously distributed one, hinting that the active region where the actual phenomena happens is much smaller than the device size.

Endurance characteristics of LSMO/SRO superlattice memory cell are shown in fig. 3(c). Memory window of the cell is found using the formula ($R_{OFF}$-$R_{ON}$)/$R_{ON}$≈$R_{OFF}$/$R_{ON}$, is nearly 14, such a huge memory margin essentially makes the device circuit capable to distinguish between the storage information 1 and 0. During the cycling there exist scattering of the resistance in

HRS, however, due to high $R_{OFF}/R_{ON}$ ratio of the present device, this kind of scattering may be admitted. It can be seen that the memory margin keeps beyond 14 times during cycling, and the cell shows little degradation even after repeated sweep cycles. Hence, the endurance measurements essentially ensure the switching between on and off states is highly controllable, reversible and reproducible. We could switch the device for about 30 – 40 times. No electrical power is required to maintain the resistance state within the given state after the device switched on or off.

Fig. 3(d) shows the threshold voltage of SET and RESET process when the device is repeatedly switched between high resistance and low resistance states. From our observation, it is evident that the $V_{SET}$ varies from 0.5 to 1 V whereas $V_{RESET}$ varies between -0.5 to -0.6 V, which indicates more variation for $V_{SET}$ probably due to the filamentary switching. *Zeng Wang et. al,* [29] reported that competition between different filaments paths can decide the formation of key conducting filaments in SET process. So it is more random than break of filaments in the reset process resulting in different variation between $V_{RESET}$ and $V_{SET}$.

Conceivable mechanisms for the resistive switching behaviors discussed until now in LSMO/SRO superlattices can be interpreted by the conducting filament model[30, 31]. We believe that the actual phenomenon happens in the top layer of the superlattice. The possible mechanism could be that under a high electrical field, oxygen ions in the top layer of the superlattice (SRO) might have migrated from the lattice positions, as a result of thermal effects and such oxygen vacancies can be seen in the form of defects. This would eventually results change in the stoichiometry and enhancement in the electronic conductivity of the superlattice. Aforementioned vacancies form local paths/filaments which might eventually switch the device to low resistance state. The change of the device from low resistance state to high resistive state is possible by rupture of filament, which essentially lead the switching of a device from a LRS to a high resistive state (HRS, or "OFF state") is labeled as RESET process. We believe that filaments in the present superlattice might have formed in the top layer as this device consists of planar structure. Finally, superlattice structure with LSMO/SRO allowed us to get stable switching. Schematic of such phenomenon is shown in Fig. 4. Moreover, as the superlattice structure merges the low dimensionality of the individual layer thicknesses with the utility of large – scale films, we could achieve unique and potentially important stable resistance switching throughout transport measurement.

In conclusion, a stable bipolar resistive switching characteristic prevails in LSMO/SRO superlattice. Merging of a superlattice structure with the low dimensionality of individual layer allowed us to attain very stable and reproducible bipolar resistive switching characteristics. We also ascertain that the memory window that is observed between ON and OFF state would indeed allow one to distinguish the information, which is a potential characteristic of present superlattice device for future data storage applications.

**Acknowledgements**

SNJ would like to thank IIT Hyderabad and K.U. Leuven Excellence financing (INPAC), the Flemish Methusalem financing and the IAP network of the Belgian Government. SNJ would also like to thank Dr. Ionela Vrejoiu for providing the superlattice sample and Dr. Eckhard Pippel and Dr. Miryam Arrdondo from Max-Planck-Institut für Mikrostrukturphysik (MPI) – Halle for the

STEM investigation of the sample.

**References:-**

1. James D. Meindl, Q. Chen, and J. A. Davis, *Science* **293**, 2044 (2001).
2. Rainer Waser and Masakazu Aono, *Nature Materials* **6**, 833 (2007).
3. G. I. Meijer, Science **319**, 1625 (2008).
4. M. J. Lee, Y. Park, D. S. Suh, E. H. Lee, S. Seo, D. C. Kim, R. Jung, B. S. Kang, S. E. Ahn, C. B. Lee, D. H. Seo, Y. K. Cha, I. K. Yoo, J. S. Kim and B. H. Park, *Adv. Mater.* **19**, 3919 (2007).
5. J. C. Bruyere and B. K. Chakraverty, *Appl. Phys. Lett.* **16**, 40 (1970).
6. K. L. Chopra, *J. Appl. Phys.* **36**, 184 (1965).
7. A. Beck, J. G. Bednorz, Ch. Gerber, C. Rossel, and D. Widmer, *Appl. Phys. Lett.* **77**, 139 (2000).
8. Alan Kalitsov, Ajeesh M. Sahadevan, S. Narayana Jammalamadaka, Gopinadhan Kalon, Charanjit S. Bhatia, Guangcheng Xiong and Hyunsoo Yang, *AIP Advances* **1**, 042158 (2011).
9. Byungjin Cho, Sunghun Song, Yongsung Ji, Tae-Wook Ki, and Takhee Lee, *Adv. Funct. Mater*, **21**, 2806 (2011)
10. Y. C. Bae, Ah Rahm Lee, Ja Bin Lee, Ja Hyun Koo, Kyung Cheol Kwon, Jea Gun Park, Hyun Sik Im, and Jin Pyo Hong, *Adv. Funct. Mater,* **22**, 709 (2012)
11. Jay A. Switzer, Rakesh V. Gudavarthy, Elizabeth A. Kulp, Guojun Mu, Zhen He, and Andrew J. Wessel, *J. Am. Chem. Soc.* **132**, 1258 (2010)

12. Michael C. Martin, G. Shirane, Y. Endoh and K. Hirota, Y. Moritomo and Y. Yokura *Phys. Rev. B.* **53** 14285 (1996).

13. J. J. Hamlin, S. Deemyad, J. S. Schilling, M. K. Jacobsen, R. S. Kumar, and A. L. Cornelius, G. Cao and J. J. Neumeier *Phys. Rev. B.* **76**, 014432 (2007).

14. 14. S. Narayana Jammalamadaka, J. Vanacken and V. V. Moshchalkov *Euro.Phys. Lett.,* **98** (2012) 17002

15. M. Ziese , I. Vrejoiu and D. Hesse *Appl. Phys. Lett.* **97** 052504 (2010)

16. M. Ziese, I. Vrejoiu, E. Pippel, P. Esquinazi, D. Hesse, E. Etz, J. Henk, A. Ernst, I. V. Maznichenko, W. Hergert and I. Merting *Phys. Rev. Lett.* **104** 167203(2010)

17. Hillebrand R., Pippel E. and Hesse D., *Phys. Status Solidi (a),* **208** 2144 (2011)

18. M. Ziese and I. Vrejoiu, *Phys. Status Solidi RRL* **7**, No. 4, 243–257 (2013)

19. Z. Q. Liu, Y. Ming, W. M. Lü, Z. Huang, X. Wang, B. M. Zhang, C. J. Li, K. Gopinadhan, S. W. Zeng, A. Annadi, Y. P. Feng, T. Venkatesan, and Ariando, *Appl. Phys. Lett.* 101, 223105 (2012).

20. C.H. Ahn, R.H. Hammond, T.H. Geballe, M.R. Beasley, J.-M. Triscone, M. Decroux, Ø. Fischer, L. Antognazza, and K. Char, Appl. Phys. Lett.**70**, 206 (1997).

21. D. Toyota, I. Ohkubo, H. Kumigashira, M. Oshima, T. Ohnishi, M. Lippmaa, M. Takizawa, A. Fujimori, K. Ono, M. Kawasaki, and H. Koinuma, Appl. Phys. Lett.**87**, 162508 (2005).

22. Priya Mahadevan, F. Aryasetiawan, A. Janotti, and T. Sasaki, Phys. Rev. B 80, 035106 (2009).

23. Hu Young Jeong, Jeong Yong Lee, Sung-Yool Choi and Jeong Won Kim *Appl. Phys. Lett.* **95**, 162108 (2009)

24. R. Müller, J. Genoe and P. Heremans *Appl. Phys. Lett.* **95**, 133509 (2009)

25. Bharti Singh, B. R. Mehta, Deepak Varandani, Govind, A. Narita, X. Feng and K. Müllen *J. Appl. Phys.* **113**, 203706 (2013).

26. A. Lampert and P. Mark, Current Injection in Solids (Academic, New York, 1970).

27. Q. Liu, W. H. Guan, S. B. Long, R. Jia, M. Liu, and J. N. Chen, *Appl. Phys. Lett.* **92**, 012117 (2008)

28. K. Jung, H. Seo, Y. Kim, H. Im, J. Hong, J. W. Park, and J. K. Lee, *Appl. Phys. Lett.* **90**, 052104 (2007).

29. Z. Wang, P. Giffin, J. McVittie, S. Wong, P. McIntyre, and Y. Nishi, *IEEE Electron Device Lett.* **28**, 14 (2007)

30. Jui-Yuan Chen, Chun-Wei Huang, Chung-Hua Chiu, Yu-Ting Huang, Su-Jien Lin, Wen-Wei Wu, and Lih-Juann Chen, *Nano Lett.*, **13** (8), 3671 (2013)

31. Seul Ji Song, Jun Yeong Seok, Jung Ho Yoon, Kyung Min Kim, Gun Hwan Kim, Min Hwan Lee & Cheol Seong Hwang *Scientific Reports* **3**, Article number: 3443 (2013)

**Figure captions:**

**Fig. 1:** (a) Schematic of device that was made using $La_{0.7}Sr_{0.3}MnO_3$ (LSMO)/ $SrRuO_3$ (SRO) superlattice. The superlattice consist of 30 layers of LSMO (2.3 nm) and SRO (3.3 nm) deposited on $SrTiO_3$ (001), as shown in the schematic. (b) HAADF-STEM micrograph of the $La_{0.7}Sr_{0.3}MnO_3$ / $SrRuO_3$ superlattices: the brown rectangles indicate the interfacial regions that are affected by intermixing.

**Fig. 2:** Bipolar resistive switching characteristics of LSMO/SRO superlattice at room temperature (300 K) . (a) I – V characteristics indicates that the current switching behavior is quite stable. (b) Shows the same graph in logarithmic plot

**Fig. 3:** (a) Current vs voltage plot in logarithmic scales of LSMO/SRO superlattice during positive voltage sweep (b) Current vs voltage plot in logarithmic scales of LSMO/SRO superlattice device during negative voltage sweep (c) Endurance characteristics of LSMO/SRO superlattice memory cell. Memory window of the cell is found using the formula ($R_{OFF}$-$R_{ON}$)/$R_{ON}$≈$R_{OFF}$/$R_{ON}$, is nearly 14, such a huge memory margin essentially make the device circuit to distinguish the storage information between 1 and 0 (d) Threshold voltage of SET and RESET process with respect to the cycle number when the device is repeatedly switched between high resistance and low resistance states

**Fig. 4:** Schematic to explain the possible mechanism for the resistive switching in LSMO/SRO superlattices. Essentially, the filaments in the present superlattice might have formed along the surface as this device consists of planar structure.

**Figures**

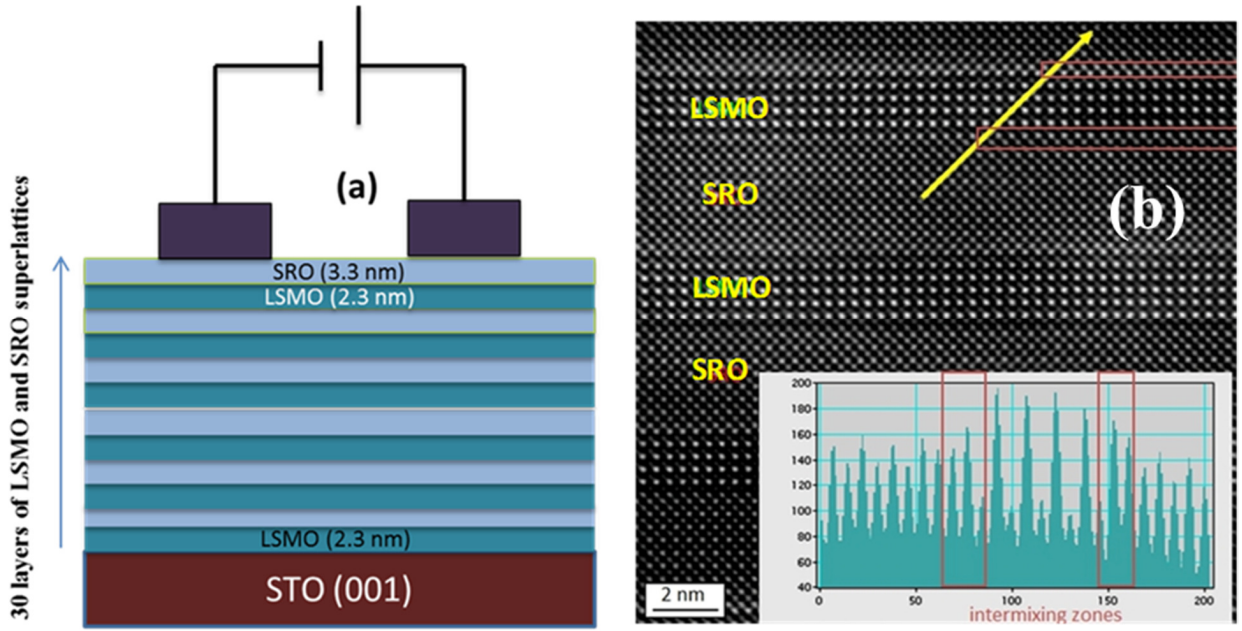

**Fig. 1: (a)** Schematic of device that was made using $La_{0.7}Sr_{0.3}MnO_3$ (LSMO)/ $SrRuO_3$ (SRO) superlattice. The superlattice consist of 30 layers of LSMO (2.3 nm) and SRO (3.3 nm) deposited on $SrTiO_3$ (001), as shown in the schematic. **(b)** HAADF-STEM micrograph of the $La_{0.7}Sr_{0.3}MnO_3$ / $SrRuO_3$ superlattices: the brown rectangles indicate the interfacial regions that are affected by intermixing.

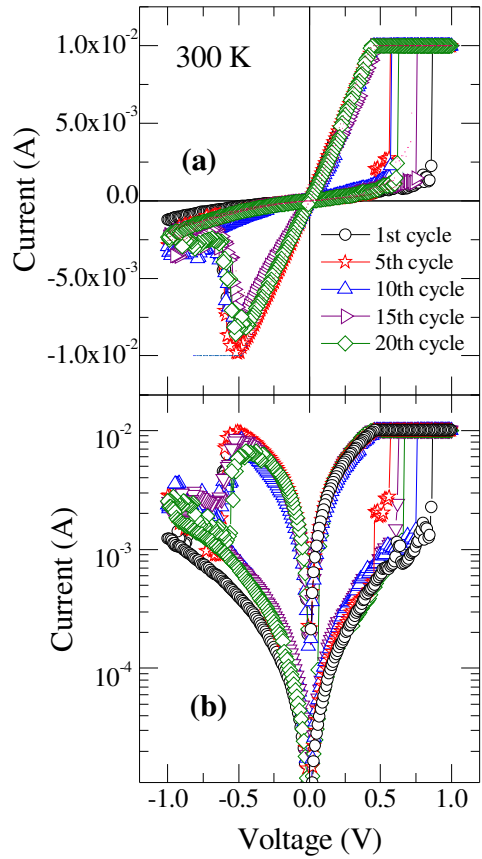

**Fig. 2:** Bipolar resistive switching characteristics of LSMO/SRO superlattice at room temperature (300 K) . **(a)** I – V characteristics indicates that the current switching behavior is quite stable. **(b)** Shows the same graph in logarithmic plot

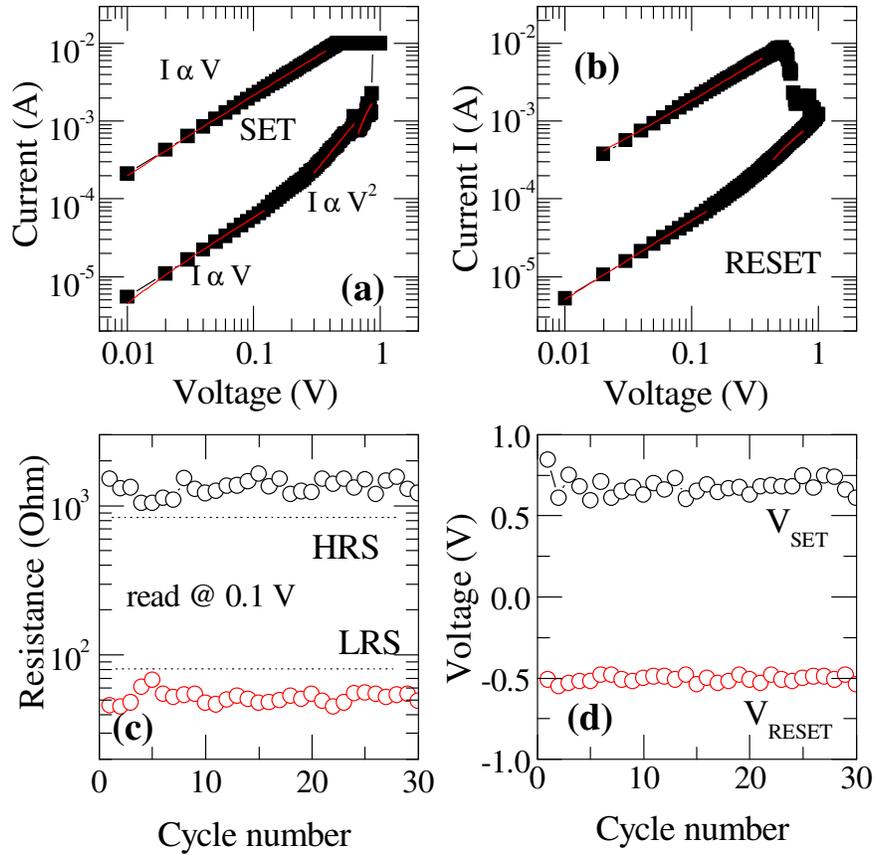

**Fig. 3: (a)** Current vs voltage plot in logarithmic scales of LSMO/SRO superlattice during positive voltage sweep **(b)** Current vs voltage plot in logarithmic scales of LSMO/SRO superlattice device during negative voltage sweep **(c)** Endurance characteristics of LSMO/SRO superlattice memory cell. Memory window of the cell is found using the formula ($R_{OFF}-R_{ON}$)/$R_{ON} \approx R_{OFF}/R_{ON}$, is nearly 14, such a huge memory margin essentially make the device circuit to distinguish the storage information between 1 and 0 **(d)** Threshold voltage of SET and RESET process with respect to the cycle number when the device is repeatedly switched between high resistance and low resistance states

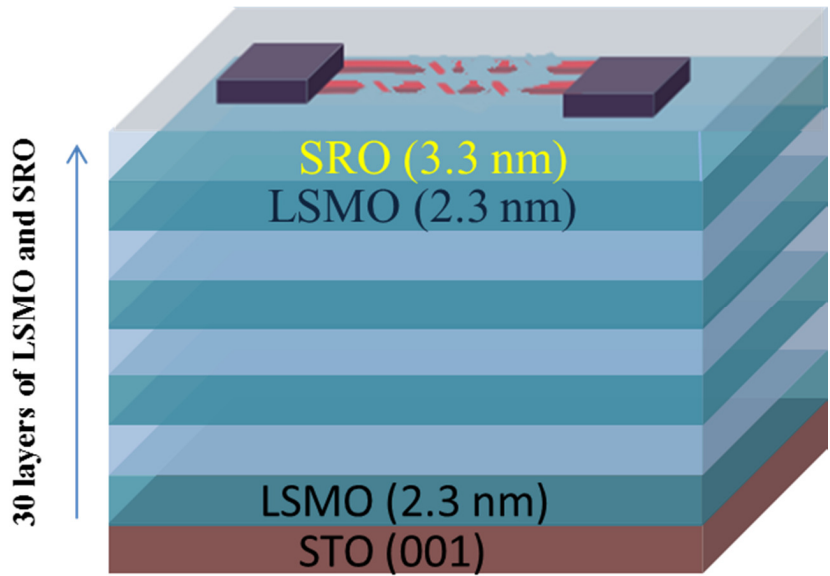

**Fig. 4:** Schematic to explain the possible mechanism for the resistive switching in LSMO/SRO superlattices. Essentially, the filaments in the present superlattice might have formed along the surface as this device consists of planar structure.